\documentclass[10pt,preprint2]{aastex}

\textheight=\baselinestretch\textheight
\ifdim\textheight>25.2cm\textheight=25.0cm\fi
\topskip\baselineskip
\maxdepth\baselineskip
\usepackage{epsfig}
\usepackage{amsmath}
\usepackage{graphicx}
\usepackage{epstopdf}
\usepackage{longtable}
\usepackage{lscape}
\usepackage{natbib}

\newbox\grsign \setbox\grsign=\hbox{$>$}
\newdimen\grdimen \grdimen=\ht\grsign
\newbox\laxbox \newbox\gaxbox
\setbox\gaxbox=\hbox{\raise.5ex\hbox{$>$}\llap
     {\lower.5ex\hbox{$\sim$}}}\ht1=\grdimen\dp1=0pt
\setbox\laxbox=\hbox{\raise.5ex\hbox{$<$}\llap
     {\lower.5ex\hbox{$\sim$}}}\ht2=\grdimen\dp2=0pt

\shorttitle{Statistical Properties of Galactic $\delta$ Scuti Stars}
\shortauthors{Chang et al.}

\begin{document}
 \title{STATISTICAL PROPERTIES OF GALACTIC $\delta$ SCUTI STARS: REVISITED}

\author
{S.-W. Chang\altaffilmark{1}, P. Protopapas\altaffilmark{2,3}, D.-W. Kim\altaffilmark{1,2,\dag} and Y.-I. Byun\altaffilmark{1}}
\altaffiltext{1}{Department of Astronomy, Yonsei University, Seoul, Korea, 120-749; seowony@galaxy.yonsei.ac.kr}
\altaffiltext{2}{Harvard Smithsonian Center for Astrophysics, Cambridge, MA 02138, USA}
\altaffiltext{3}{Institute for Applied Computational Science,  School of Engineering and Applied Sciences, Harvard, Cambridge, MA 02138, USA}
\altaffiltext{\dag}{Present address: Max Planck Institute for Astronomy, K\"{o}nigstuhl 17, D-69117 Heidelberg, Germany; kim@mpia-hd.mpg.de}


\begin{abstract}
We present statistical characteristics of 1,578 $\delta$ Scuti stars including nearby field stars and cluster member stars within the Milky Way.  We obtained 46\% of these stars (718 stars) from the works done by Rodr\'{i}guez and collected the remaining 54\% stars (860 stars) from other literatures.  We updated the entries with the latest information of sky coordinate, color, rotational velocity, spectral type, period, amplitude and binarity.  The majority of our sample are well characterized in terms of typical period range ($0.02-0.25$ days), pulsation amplitudes ($<$0.5 mag) and spectral types (A--F type).  Given this list of $\delta$ Scuti stars, we examined relations between their physical properties (i.e., periods, amplitudes, spectral types and rotational velocities) for field stars and cluster members, and confirmed that the correlations of properties are not significantly different from those reported in the Rodr\'{i}guez's works.  All the $\delta$ Scuti stars are cross-matched with several X-ray and UV catalogs, resulting in 27 X-ray and 41 UV-only counterparts.  These counterparts are interesting targets for further study because of their rarity and uniqueness in showing  $\delta$ Scuti-type variability and X-ray/UV emission at the same time. The compiled catalog can be accessed through the web interface (\url{http://stardb.yonsei.ac.kr/DeltaScuti)}.
\end{abstract}


\section{Introduction}
\label{sec:introduction}
Observation of pulsating variable stars is a unique approach to study  internal stellar structures and evolutionary status across a broad range of stellar types \citep{gau96}. Typical examples of pulsating variables are hydrogen-rich DA white dwarfs, Cepheids, RR Lyraes, $\delta$ Scutis and $\gamma$ Doradus stars.  In particular, $\delta$ Scuti stars (hereinafter, $\delta$ Sct stars) have attracted much attention in recent years because of their great number of radial and non-radial pulsation modes, driven by the $\kappa$ mechanism that mostly works in the He II ionization zone (e.g., \citealp{gau95}; \citealp{bre98}; \citealp{rod01}; \citealp{bre02}; \citealp{bre05}).  They are located at the lower part of the instability strip occupied by main sequence (MS), pre-MS and more evolved stars, and characterized by relatively short-period and low-amplitude variability.  It is well known that periods of classical $\delta$ Sct stars are in the range of 0.02--0.25 days, with amplitudes between 0.003 and 0.9 magnitudes in $V$ band\footnote{The pulsation amplitude has an intrinsic dispersion caused by the range of periods and amplitudes, and so we adopt the amplitude range that is defined in the GCVS variability classification scheme (see \url{http://www.sai.msu.su/gcvs/gcvs/iii/vartype.txt)}.}.  Note that the lower side of these amplitude values is just an observational limit of the ground-based surveys.  Kepler have detected a large number of $\delta$ Sct stars whose amplitudes are lower than 0.003 magnitude (e.g., \citealt{uyt11, bal11b}).  Also classical $\delta$ Sct stars have masses between 1.6 and 2.4 $M_\odot$ for near solar-metallicity stars, and between 1.0 and 1.3 $M_\odot$ for metal-poor ($-1.5<$ [Fe/H] $<-1.0$) stars \citep{mcn11}.  There are numbers of detailed reviews describing $\delta$ Sct stars and associated issues (e.g., see \citealp{bre00,rod01}; \citealt{lam00} and references therein).

In order to update statistical view of pulsational and physical characteristics of galactic $\delta$ Sct stars, we collected a large sample of $\delta$ Sct stars from (1) the catalog of $\delta$ Sct-type variables (\citealt{rod00a}, hereinafter R2000), (2) the catalog of SX Phe type variables in globular clusters \citep{rod00b} and (3) other individual findings after R2000 (e.g., \citealp{hen01,dal02,sok02,rod03,ber04,cha04,hen02,esc05,mar05,zha06,chr07,jeo07,pen07,har08,pig09,sok09,soy09}).  Note that we compiled $\delta$ Sct stars that appeared in the literatures published before 2011.

Not surprisingly, the quantitative increase of $\delta$ Sct stars is due to an increasing number of long-term variability observations that surveyed a large fraction of sky.  MACHO \citep{alc00} and OGLE \citep{uda97} monitored the galactic bulge for several years, and discovered a number of pulsating variables including $\delta$ Sct-type variables.  Other wide-field all-sky surveys also detected more than 500 $\delta$ Sct stars (ROTSE: \citealp{bla03}, \citealp{jin03}; ASAS: \citealp{poj06}; TAOS: \citealp{kim10}).  In order to study the low-amplitude oscillations of the pulsators in the cluster, several individual observations monitored nearby open clusters with the age range from 0.012 to 2.8 Gyrs (e.g., \citealp{fre01,are05,mar05,kan07, and09,jeo08,jeo09a,jeo09b}) and found nearly one hundred $\delta$ Sct stars.  In addition, a large number of metal-poor $\delta$ Sct stars (SX Phe stars) have been  discovered even in the central regions of globular clusters \citep{pyc01,bru01,jeo01,jeo03,maz03,jeo04,kop05,ole05,kop07,are08,are10} using improved photometry techniques (e.g., difference image analysis).  GCVS \citep{sam09} catalog also provides another $\sim$400 pulsating stars designated as DSCT or DSCTC.  DSCTs are variables of the $\delta$ Sct-type that are close to the SX Phe-types, while DSCTCs are low-amplitude group of $\delta$ Sct variables ($\Delta$V $<$ 0.1 mag).  From the above literatures, we include all pulsating stars mentioned as $\delta$ Sct stars for this work.

In Section 2, we describe procedures for selecting typical $\delta$ Sct stars and brief information of the catalog format.  In Section 3, we present statistical distributions of physical properties and their relationships.  In Section 4, we present lists of X-ray/UV counterparts and their characteristics.  A concluding summary is presented in the last section.\\

\section{Catalog compilation}
\label{sec:catalog compilation}
We first collected the largest possible samples of field and cluster $\delta$ Sct stars that were identified by previous studies as typical $\delta$ Sct-type.  The catalog contains subclasses of $\delta$ Sct stars showing clear differences in the metal abundances due to diffusion and other processes \citep{rod01} and thus representative of chemically peculiar $\delta$ Sct star's diversity such as:

\begin{itemize}
\item The group of $\lambda$ Bootis ($\lambda$ Boo) type stars that are defined as metal-poor (except C, N, O and S elements which have a solar abundance) Population I objects \citep{pau97} 
\item The classical and evolved ($\rho$ Pup and $\delta$ Del) metallic-line stars (Am stars), which are characterized by an underabundance of C, N, O, Ca and Sc, as well as by an overabundance of the Fe group and heavier elements \citep{har11}
\item Metal-poor SX Phe stars of Population II and old disk population
\end{itemize}

We also include the coolest subgroup of Ap stars known as the rapidly oscillating Ap (roAp) stars. Among them, $\delta$ Sct- or $\gamma$ Dor-type pulsations are clearly present \citep{bal11a}. The final group is a small number of pulsating pre-main-sequence (PMS) stars. PMS pulsators with masses between 1.5 and 4 $M_\odot$ are known to have the consistent spectral types, luminosities and pulsation modes with classical $\delta$ Sct stars \citep{zwi08}. 

We removed all stars that turned out not to be $\delta$ Sct stars by later studies.  These stars are either W UMa and RR Lyrae stars, or stars showing no evidence of periodicity in their light curves (e.g., V1241 Tau: \citealp{are04}).  Finally, our catalog contains 1,578 $\delta$ Sct stars within our galaxy, which provides relatively more complete and up-to-date entries than the previous works of similar nature (R2000 and \citealt{rod00b}).  The number of pre-existing and newly listed $\delta$ Sct stars is summarized in Table \ref{tab:Table1}.  The total number of $\delta$ Sct stars, including 1,282 field stars and 296 cluster member stars, is increased by factor of two in comparison with the existing catalogs.
    
In order to eliminate duplicated entries and to extract their additional physical parameters, all $\delta$ Sct stars in the catalog were crossmatched with the VizieR CDS database \citep{gen00} either by their designations (i.e., HD, HIP or GCVS designation) or by a radial search.  Table \ref{tab:Table2} shows the catalogs crossmatched with the $\delta$ Sct stars.\\ 

\begin{table*}
\caption{Number of $\delta$ Sct stars \label{tab:Table1}}
\begin{center}
\begin{tabular}{cccc}
\tableline\tableline
 Sources & Existing catalogs$^a$ & This work$^b$ & References \\
\tableline
 
Hippacros & 78 & 78  & \citet{per97} \\          
OGLE$^c$  & 52 & 52  & \citet{uda97} \\
MACHO$^d$ & 81 & 81  & \citet{alc00} \\
ROTSE$^e$ & -  & 4   & \citet{jin03}    \\
ASAS$^f$  & -  & 525 & \citet{poj06} \\
TAOS$^g$  & -  & 41  & \citet{kim10}  \\
GCVS$^h$  & 294 & 419 & \citet{sam09} \\

& & & \\
Open clusters & 64 & 92 & List A$^i$ \\           
Globular clusters & 123 & 204 & List B$^j$   \\   
Miscellaneous & 26 & 82 & Individual papers \\
\tableline
Total & 718 & 1,578 &  \\
 \tableline
 \end{tabular}
 \end{center}
 \begin{flushleft}
  $^a$ The catalogs compiled by R2000 and \citet{rod00b}. \\
  $^b$ This number (column 3) is inclusive of both pre-existing (column 2) and newly listed $\delta$ Sct stars.\\
  $^c$ Optical Gravitational Lensing Experiment. \\
  $^d$ MAssive Compact Halo Object. \\ 
  $^e$ Robotic Optical Transient Search Experiment. \\
  $^f$ All-Sky Automated Survey. \\
  $^g$ Taiwan-American Occultation Survey. \\
  $^h$ General Catalog of the Variable Stars. \\
  $^i$ List A: $\alpha$ Persei, Pleiades, Hyades, Praesepe, Melott 71, NGC 2682, NGC 3496, NGC 5999, NGC 6134, NGC 6882, NGC 7062, NGC 7245, NGC 7654, IC 4756 \\
  $^j$ List B: $\omega$ Cen, 47 Tuc, IC4499, M3, M4, M5, M13, M15, M53, M55, M56, M68, M71, M92, NGC 288, NGC 3201, NGC 4372, NGC 5053, NGC 5466, NGC 5897, NGC 6362, NGC 6366, NGC 6397, NGC 6752, Ru 106 \\
  \end{flushleft}
\end{table*}

\begin{table*}
 \begin{center}
 \caption{Catalogs used to extract additional parameters \label{tab:Table2}}
 \begin{tabular}{cc}
\tableline\tableline
 Catalog Name & References \\
\tableline

Henry Draper Catalogue and Extension & \citet{can93} \\ 
HDE Charts: position, proper motions & \citet{nes95} \\
The Hipparcos and Tycho Catalogue & \citet{per97} \\
Catalog of Projected Rotational Velocities & \citet{gle00} \\
The Catalogue of Components of Double and Multiple & \citet{dom02}\\
NOMAD Catalog & \citet{zac05} \\
The Guide Star Catalog, Version 2.3.2 & \citet{las06} \\
Rotational velocities of A-type stars. III. Velocity dispersions & \citet{roy07} \\
Rotational Velocity Determinations for 118 $\delta$ Sct Variables & \citet{bus08} \\
General Catalogue of Variable Stars & \citet{sam09} \\
All-Sky Compiled Catalogue of 2.5 million stars & \citet{kha09} \\
AAVSO International Variable Star Index VSX & \citet{wat09} \\
The Washington Visual Double Star Catalog & \citet{mas09} \\
Catalogue of Stellar Spectral Classifications & \citet{ski09} \\

 \tableline
 \end{tabular}
 \end{center}
\end{table*}

\begin{table*}
 \begin{center}
 \caption{New catalog of $\delta$ Sct stars \label{tab:Table3}}
 \begin{tabular}{ccccccccccccc}
\tableline\tableline
 ID & RA & Dec & $V$ & $B$ & Period\tablenotemark{a} & $\Delta$$V$\tablenotemark{b} & $v$sin$i$ & SpType & S/P\tablenotemark{c} & B/M\tablenotemark{d} & Type\tablenotemark{e} & GCVS \\
    & (h:m:s) & (d:m:s) & (mag) & (mag) & (days) & (mag) & (km/s) &  & & & & (name)\\
\tableline
1 & 00 00 53 & $+$62 25 15 & 15.80 &      & 	   & 0.40  &  &
    &  &  0 & MWF & V0878 Cas \\ 
2 & 00 01 16 & $-$60 37 00 & 9.93 & 10.33 & 0.1221 & 0.35 &  & A8V & S & 0 & MWF & \\ 
3 & 00 01 16 & $+$06 47 29 & 7.23 & 7.62 & 0.1652 & 0.04 & 135 & F0 & P & 0 & MWF & DR Psc \\ 
4 & 00 04 00 & $+$12 08 45 & 7.26 & 7.62 & 0.1701 & 0.06 & 74 & F0III & S & 1 & MWF & NN Peg\\ 
5 & 00 04 12 & $-$20 55 06 & 11.66 &  & 0.1790 & 0.17 &  & &  & 0 & MWF & \\ 
6 & 00 05 54 & $+$11 28 18 & 13.59 &  & 0.1588 & 0.52 &  &  &  & 0 & MWF &  \\
\tableline
 \end{tabular}
\tablecomments{Table \ref{tab:Table3} is published in its entirety in the electronic edition of {\it Astronomical Journal}. A portion is shown here for guidance regarding its form and content.}
\tablenotetext{a}{Period corresponds to the dominant pulsation mode}
\tablenotetext{b}{Peak-to-peak magnitude}
\tablenotetext{c}{Spectroscopic spectral type = S, Photometric spectral type = P}
\tablenotetext{d}{Single stars = 0, Binary or multiple stars = 1}
\tablenotetext{e}{Milky way field stars = MWF, Open cluster member stars = OC, Globular cluster member stars = GC}
 \end{center}
 \end{table*}

\subsection{Catalog description}
\label{sec:catalog description}
 Table \ref{tab:Table3} presents a part of our catalog of $\delta$ Sct stars that lists the IDs, coordinates (J2000.0), mean magnitudes, periods, amplitudes, rotational velocities, binarities and spectral types with comments.  In addition, we specified the membership for each $\delta$ Sct star by three groups: milky way field stars (MWF), globular cluster member stars (GC) and open cluster member stars (OC), respectively.

{\underline{\it ID}}:  The numbering of stars in the catalog is the order of increasing right ascension, labeled as DS1--DS1578. We retain the previous numbering system from R2000 and \citet{rod00b} which will only appear in the electronic version of the journal.

{\underline{\it Magnitudes}}:  Our catalog lists the apparent magnitude $m_V$ (or $m_B$) that is denoted $V$ (or $B$) for all objects, and at least includes $V$ magnitude if available from one of the catalogs listed in Table \ref{tab:Table2}.  In the OGLE and ASAS surveys, $V$ magnitude is calculated as: $V$ = $min(V)$ (minimum magnitude) + $\Delta V$/2 (half-amplitude of variability).  Except for 17 stars, all $\delta$ Sct stars in the catalog have $V$ magnitude.

{\underline{\it Period and amplitude}}:  Many light curves of $\delta$ Sct stars show signs of multiperiodicity due to the simultaneously excited radial and/or non-radial modes (e.g., \citealp{kis02}).  However, we only list the periods and amplitudes which correspond to the dominant primary periodicity in the period search.  Similar to previous definition in R2000, amplitudes correspond to the full amplitudes ($\Delta$$V$) of periodic variations in the $V$ band.  If reliable amplitude estimation is not available, we defined the amplitude as the difference between the maximum and minimum $V$ magnitude.

{\underline{\it Rotational velocity}}:  We used three dedicated catalogs \citep{gle00,roy07,bus08} to extract the projected rotational velocities ($v$sin$i$) and found $v$sin$i$ for about 10$\%$ of the stars.  Due to the differences between the methods for measuring stellar rotation, these catalogs often present slightly different values for $v$sin$i$. \citet{roy07} noticed that there is a systematic deviation of each method and used their Fourier transform scale of $v$sin$i$. On the other hand, \citet{bus08} examined the relation between their measured $v$sin$i$ values and the average $v$sin$i$ values from the published literatures, and concluded that there is no systematic offset within the error of each measurement. Here we simply selected $v$sin$i$ values from the most recently published data, regardless of measuring methods.

{\underline{\it Spectral type}}:  We compiled the spectral type information primarily based on the catalog by \citet{ski09}, which includes only spectral types determined from spectra (i.e., line and band strengths or ratios).  The spectral types are expressed in the MK notation only for $\delta$ Sct stars.  We do not list the individual spectral types of binary companions.  When the spectroscopic spectral types are not available, we took the spectral types determined from photometry or inferred from broadband colors.  In our catalog, we found spectroscopic spectral types (denoted as `S' in the catalog) for 15\% of the stars and photometric estimates (P) for 9\% of the stars, respectively.

{\underline{\it Binarity}}:  It is known that many $\delta$ Sct stars are components of binary or multiple star systems (see Fig.2 of \citealp{zho10}), and that their pulsation properties may be modified by binarity \citep{lam00}.  We searched the double star catalogs \citep{nes95,dom02,mas09} and found information on binarity for 141 stars, which are indicated by flag `1'.\\  

\section{Statistical properties of $\delta$ Sct stars}
\label{sec:statistical properties}

\subsection{Histograms of the pulsational properties and the physical properties}
\label{sec:HistogramsOfProperties}
In this section we present histograms which describe the distribution of each parameter (magnitude, period, amplitude, $v$sin$i$ and spectral type) for all $\delta$ Sct stars according to their membership groups (i.e., MWF, GC or OC) defined in the previous section.  The members of binary or multiple star systems are excluded from the histograms to reduce potential contaminations on the parameters.

\begin{figure}[t]
\begin{center}
 \includegraphics[width=0.48\textwidth, angle=0]{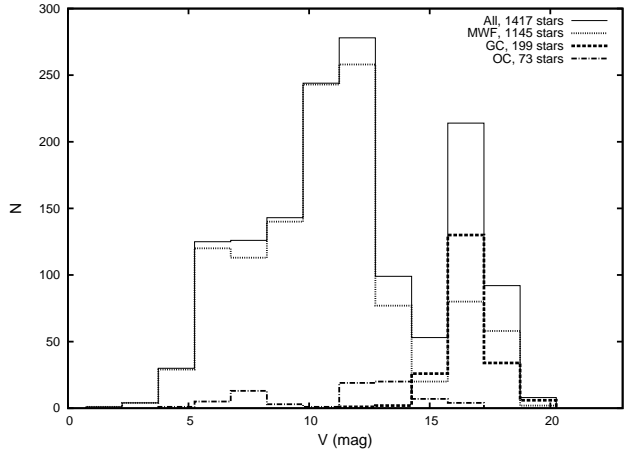}    
 \caption{Magnitude histogram of the $\delta$ Sct stars. Each histogram represents the number of stars from the field regions (dotted line), the globular clusters (dashed line), the open clusters (dot-dashed line) and the combined entries (solid line).}
 \label{fig:Fig1}
\end{center}
\end{figure}

Figure \ref{fig:Fig1} shows the histogram of the $V$ magnitudes for 1,417 $\delta$ Sct stars.  In the magnitude distribution of field $\delta$ Sct stars, two separated peaks appear as a signature of the observational selection effect caused by the different $V$ magnitude range of different variability surveys.  For example, the $\delta$ Sct stars between 16 and 20 magnitude were detected by the MACHO and OGLE surveys, while many bright stars (6 $< V <$ 10) were observed by the Hipparcos survey.  Because the absolute $V$ magnitude of classical $\delta$ Sct stars is relatively faint ($M_{V}$ $\sim$ $1-2.5$), our samples of galactic $\delta$ Sct stars generally  have apparent magnitudes brighter than $V$ = 20 \citep{alc00}.  Thus detection of distant $\delta$ Sct stars at the faint end of the histogram requires time-consuming observations, where detection efficiency decreases.  

\begin{figure}[t]
\begin{center}
 \includegraphics[width=0.48\textwidth, angle=0]{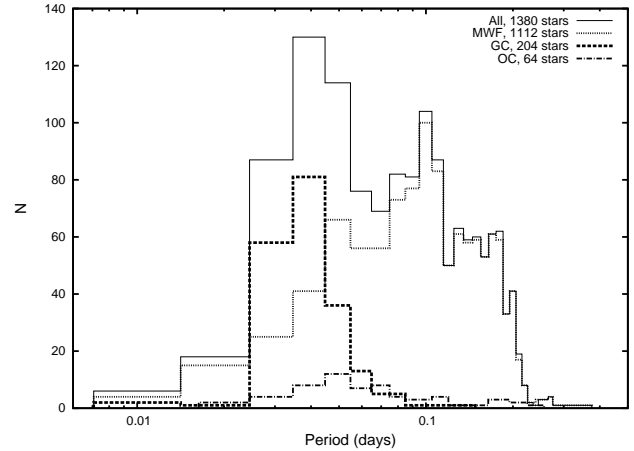}
 \caption{Period histogram of the $\delta$ Sct stars. The representation of the lines is same as Fig. 1, but the x-axis is in logarithmic scale. The dominant periods are relatively short and range from 0.02 to 0.25 days, which is a typical characteristics of $\delta$ Sct stars.}
 \label{fig:Fig2}
\end{center}
\end{figure}

Figure \ref{fig:Fig2} shows that $\delta$ Sct stars are indeed short-period variables, with individual periods lying in the range from 0.02 to 0.3 days.  In general, the period range of $\delta$ Sct stars is physically restricted between 0.02 and 0.25 days \citep{bre00}.  Our catalog includes the $\delta$ Sct stars with the shortest periods reported to date.  There are two field stars ($\sim$0.018 days) from the TAOS 2-year data \citep{kim10} and two SX Phe stars ($\sim$0.017 days) in the globular cluster $\omega$ Centauri \citep{ole05}.  As mentioned by \citet{rod01}, pulsating variables with periods between 0.25 and 0.3 days may need to be classified as the evolved Population I $\delta$ Sct or the Population II RRc (or $\gamma$ Dor).  Except for the binaries or multiple systems, nine stars belong to this period range.  One interesting object among them is UY Cam which was originally regarded as RR Lyrae type. Because of its long period (0.267 days), low metallicity (Z = 0.0037), high luminosity (M$_{V}$ = $-0.2$ mag) and low gravity (log $g$ = 3.46), this high-amplitude $\delta$ Sct star has shared physical characteristics with not only the SX Phe stars but also the RR Lyrae stars (see \citealt{zho03} and references therein).  Our catalog also includes two stars with periods longer than 0.3 days.  V4063 Sgr (=HD 185969) with a period of 0.361 days has the longest period among any known $\delta$ Sct stars \citep{mci78}, and BZ Boo (=HD 118743) has a period of 8 to 10 hours \citep{jac72}.  Further observations of these two stars are required to verify these periods and to confirm if they are $\delta$ Sct-type pulsators.

In contrast to the MWF, most SX Phe stars in globular clusters are short-period pulsators with periods less than 0.1 days.  Recent observations show that the low metal abundance seems to lead to shorter pulsation period (see Fig. 7 of \citealt{rod00b} and Fig. 1 of \citealt{mcn97}).  According to theoretical models for the evolution of stars with the low metal abundance, both fundamental ($\Pi_0$) and first-overtone ($\Pi_1$) modes are pulsationally unstable around log $\Pi_0$ = $-1.0$ \citep{tem02}.  
It shows that metal-poor stars become pulsationally unstable states mostly at periods shorter than 0.1 days.  On the other hand, some SX Phe stars with longer periods (P $>$ 0.1 days) can be explained by post-main-sequence evolution (e.g., \citealt{bru01}).

\begin{figure}[t]
\begin{center}
 \includegraphics[width=0.48\textwidth, angle=0]{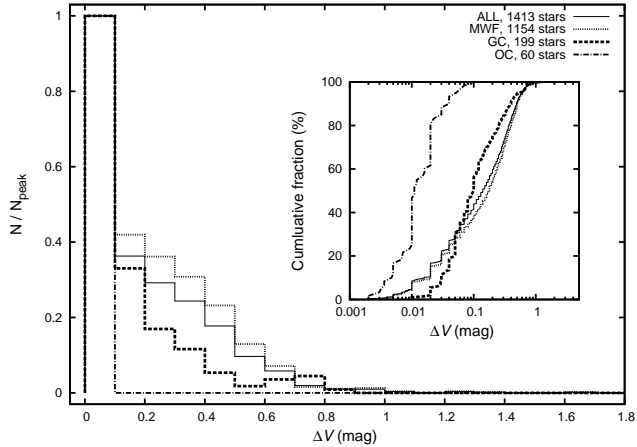}
 \caption{Normalized amplitude histogram of the $\delta$ Sct stars.  All the histograms are normalized to their peaks. The representation of the lines is same as Fig. 1, but the histogram bin size is 0$.^{m}$1. The small inserted plot shows the cumulative distribution of the amplitudes. About 40$\%$ of the $\delta$ Sct stars have pulsation amplitudes smaller than 0$.^{m}$1.}
 \label{fig:Fig3}
\end{center}
\end{figure}

Figure \ref{fig:Fig3} shows both histograms and cumulative distributions of the amplitude of $\delta$ Sct stars.  For comparison, the histograms are normalized to have a maximum value of unity.  The amplitudes are in the range from 0.002 and 1.69 mags in the $V$ band\footnote{Amplitudes in $B$ band are relatively lower than those in $V$ band.}.  Historically, on the basis of their pulsation amplitudes, the $\delta$ Sct stars are divided into low-amplitude $\delta$ Sct stars (LADS) and high-amplitude $\delta$ Sct stars (HADS).  \citet{sol97} adopted a value of $\Delta$$V$ = 0$.^{m}$1 as criterion to distinguish LADS from HADS ($\Delta$$V$ $\geq$ 0$.^{m}$3).  This amplitude difference is substantially related with their pulsation modes and evolutionary states.  Most of the LADS are on or close to the main sequence, and pulsate in non-radial p-modes, whereas HADS tend to be more evolved than the LADS, and typically pulsate in low-order radial p-modes \citep{bre00, alc00}.  Other researchers have suggested that the separation in amplitude is due to difference in rotational velocities between two groups (see Section 3.2 for details). 
In both young and intermediate age (0.012--2.8 Gyrs) open clusters, $\delta$ Sct stars tend to show only very low-amplitudes (0.002--0.1 mags), while those in field and globular clusters tend to have large range of amplitudes through the two amplitude groups.  The cumulative distributions represent that low-, medium-, and high-amplitude $\delta$ Sct stars discovered in globular clusters are in the ratio of 3:1:1, and those in field are in the ratio of 1.3:1:1.

\begin{figure}[t]
\begin{center}
 \includegraphics[width=0.48\textwidth, angle=0]{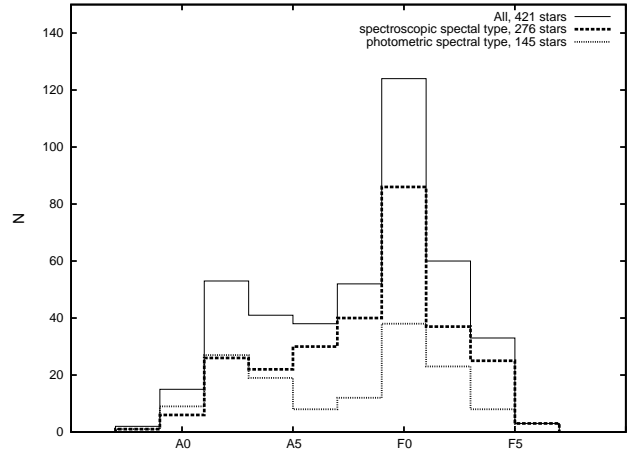}
 \caption{Spectral type histogram of the $\delta$ Sct stars. Generally there is good agreement between the overall distribution of the spectral types obtained from spectroscopic methods (dashed line) and photometric methods (dotted line).}
 \label{fig:Fig4}
\end{center}
\end{figure}

Spectral types of $\delta$ Sct stars are given in Figure \ref{fig:Fig4}.  Only 421 stars have either spectroscopic and/or photometric spectral types and their spectral types are between early A and late F type.  Although $\delta$ Sct stars generally have spectral types ranging from about A2 to F2 \citep{bre00}, our catalog includes some stars which lie outside the empirical instability strip.  Many studies have tried to obtain more accurate constraints of the $\delta$ Sct instability strip close to observed location and shape.  The blue edge of the $\delta$ Sct instability strip is theoretically well constrained \citep[see][]{pam00}, whereas the red edge is rather complicated and has a large range of possibilities for the slope and shape.  Interestingly our sample includes a few blue outliers that span a range of spectral types from A0--A2.  As suggested by \citet{sch93}, the blue edge of the instability strip may need to be extended to include the early A type stars.  On the other hand, for the low-temperature stars close to the red edge, we have to consider the coupling effect between convection and oscillation together with the turbulent viscosity.  \citet{xio01} calculated non-adiabatic oscillations for stars in the mass range $1.4-3.0$ $M_\odot$, and matched the empirical red edge very well. According to \citet{sch82}, this mass range is consistent with the spectral type of $\delta$ Sct stars between A0V and F5V.  But despite of $\delta$ Sct-like pulsation nature, about 18 red outliers actually have spectroscopic/photometric spectral types later than F5, and thus relatively cool temperatures (e.g., VX Hya (F6) and DE Lac (kF3hF7)\footnote{This star's spectral type is described as kF3hF7.  The strength of the calcium II K absorption line and the Balmer lines are more like those of an F3 star and F7 star, respectively.}).  Further spectroscopic investigations are needed in order to remove non-$\delta$ Sct stars around the blue and red edge of instability strip. 

\begin{figure}[t]
\begin{center}
 \includegraphics[width=0.48\textwidth,angle=0]{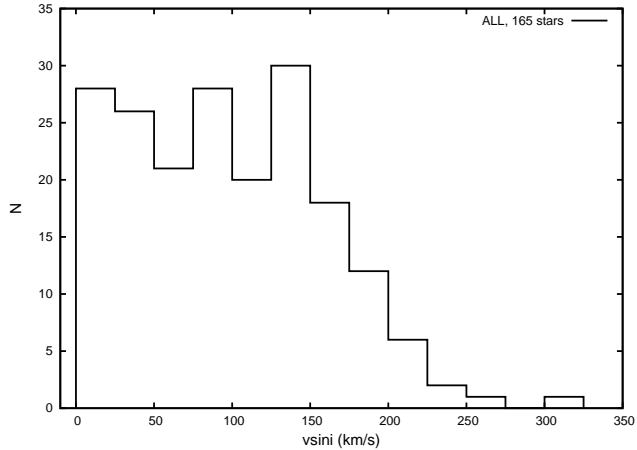}
 \caption{Projected rotational velocity ($v$sin$i$) distribution of the $\delta$ Sct stars. The number of stars rapidly decreases above the rotational velocity of 150 km/s as shown in the histogram.}
 \label{fig:Fig5}
\end{center}
\end{figure}

Figure \ref{fig:Fig5} shows the distribution of projected rotational velocities.  The distribution of rotational velocity is almost uniform for velocities smaller than 150 km/s, and  extends to 300 km/s which is 70$\%$ of break-up velocity for normal A type stars \citep{abt95}. Therefore the break-up velocity is not a limiting factor for the rotational velocities of our samples.  The broad range of rotation rates seen in $\delta$ Sct stars are in marked contrast to those of RR Lyrae stars, which have an upper limit for $v$sin$i$ of 10 km/s (Peterson, Carney $\&$ Latham 1996).\\

\subsection{Relationships between the pulsational properties and the physical properties}
\label{sec:relationships}
In this section we focus on the relationships between physical properties (spectral types, periods, rotational velocities and amplitudes) of the $\delta$ Sct stars.  Similar efforts have been made previously.  For example, \citet{ant81} showed the amplitude-period-luminosity relation for low-amplitude $\delta$ Sct stars.  Also \citet{sua02} found a significant correlation between the oscillation amplitude and rotational velocity for $\delta$ Sct stars in open clusters.  As in the previous sections, we again removed the known binaries from all the relations.

\begin{figure}[t]
\begin{center}
 \includegraphics[width=0.48\textwidth,angle=0]{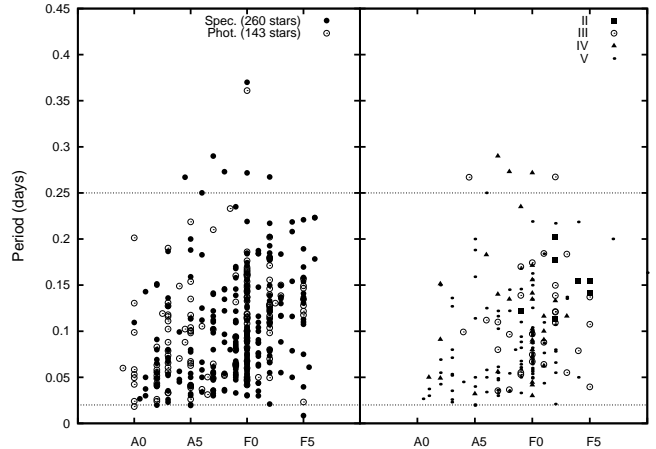}
 \caption{Relation between spectral type and period for the $\delta$ Sct stars. In the left panel, distribution between two parameters is shown for two groups of $\delta$ Sct stars either spectroscopic (dots) or photometric (open circles) spectral types. In both cases, the early type stars tend to have shorter periods than the late types. In the right panel, we indicate the same distribution with luminosity classes II--V by different symbols.}
 \label{fig:Fig6}
\end{center}
\end{figure}

As shown in Figure \ref{fig:Fig6}, there is a weak but certain relation between spectral type and period. The early type stars tend to have shorter periods than the late types. This tendency can be explained as either an evolutionary effect or observational selection effect \citep{rod00a}. In both cases, other stellar parameters have to be taken into account for deriving more reliable relation. 

\begin{figure}[!tbp]
\begin{center}
 \includegraphics[width=0.48\textwidth,angle=0]{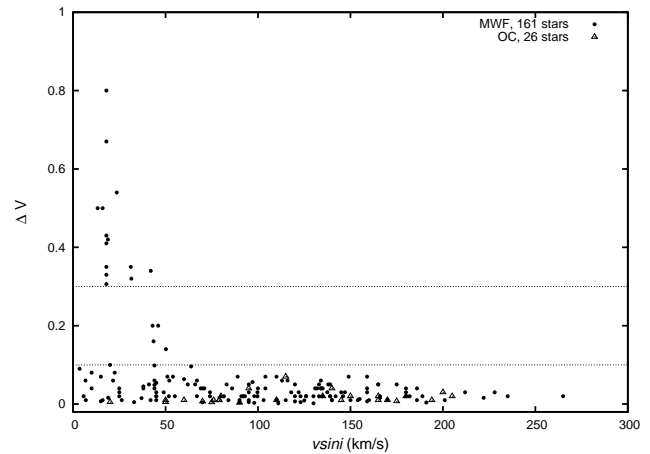}
 \caption{Relation between projected rotational velocity and full amplitude for the $\delta$ Sct stars. Field stars and OC member stars are indicated with dots and triangles, respectively. 
 HADS are pulsators with both high-amplitude magnitude variations ($\Delta$$V$ $\ge$ 0$.^{m}$3) and slow rotational velocities ($v$sin$i$ $<$ 30 km/s). In contrast, LADS have relatively rapid rotational velocities and low-amplitude variations  ($\Delta$$V$ $\le$ 0$.^{m}$1).}
 \label{fig:Fig7}
\end{center}
\end{figure}

\begin{figure}[t]
\begin{center}
 \includegraphics[width=0.48\textwidth,angle=0]{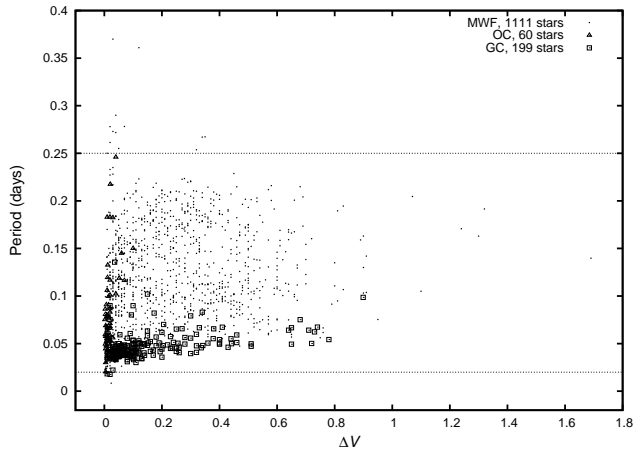}
 \caption{Relation between full amplitude and period for the $\delta$ Sct stars. Almost all MWF stars (small dots) are uniformly distributed between OC (open triangles) and GC (open squares) member stars. On the other hand, the two distributions of cluster member stars are remarkably different from the distribution of MWF stars.}
 \label{fig:Fig8}
\end{center}
\end{figure}

Figure \ref{fig:Fig7} shows a clear distinction between HADS and LADS. As mentioned in Section 3.1, $\delta$ Sct stars with large amplitudes (at least $\Delta$$V$ $\geq$ 0$.^{m}$3) are regarded as HADS that rotate slowly with $v$sin$i$ less than 30 km/s \citep{bre07b}. On the other hand, LADS ($\Delta$$V$ $\leq$ 0$.^{m}$1) have a much greater range of $v$sin$i$ including rapid rotational velocities (5.7 $\sim$ 306 km/s). Also several $\delta$ Sct stars in the open clusters show moderate or fast rotational velocities (20 $\sim$ 205 km/s), which is consistent with $v$sin$i$ values of LADS groups (\citealt{mol09, rod00a}).  A similar trend was already found for 68 $\delta$ Sct stars by \citet{sol97}, who showed that HADS tend to have lower rotation velocities, while LADS have a broader distribution in $v$sin$i$.  This empirical relation suggests that stellar rotation plays an important role in determining the size of the amplitudes of radial and non-radial modes \citep{bre07a}.  Some $\delta$ Sct stars are known to have intermediate amplitudes between HADS and LADS (0$.^{m}$1 $<$ $\Delta$$V$ $<$ 0$.^{m}$3). These stars are responsible for astrophysical connection between HADS and LADS groups.  For example, \citet{bre07b} argued that EE Cam belongs to this transition population, and also \citet{hin09} found that V873 Her, medium-amplitude $\delta$ Sct (MADS), has a similar properties to EE Cam as transition population. Other suspected transition stars (V0645 Her, V1162 Ori, V2109 Cyg and DX Cet) also show amplitudes between 0$.^{m}$1 and 0$.^{m}$2. 

Figure \ref{fig:Fig8} tells us that there is no relation between amplitudes and periods for the entire field stars, while the distributions of cluster member stars are remarkably different.  The latter stars show that long-period $\delta$ Sct stars seem to have more large amplitudes than short-period ones.  From the amplitude-temperature-period relation, as \citet{sol97} pointed out, this relation can be explained by one hypothesis that the large amplitude stars are more evolved than the low amplitude $\delta$ Sct stars. \\ 

\section{X-ray and UV counterparts of the $\delta$ Sct stars}
\label{sec:X-ray and UV counterparts}
We cross-matched our catalog with several X-ray and UV catalogs in Table \ref{tab:Table4}, and found 27 X-ray and 41 UV-only counterparts, respectively.  It is known that some binary stars such as Algol-type binaries show X-ray emission (e.g., \citealp{mcg96,ste01,che06}).  In addition, X-ray/EUV emission is known to be a properties of hot white dwarfs \citep{mar97a,mar97b}.  The general properties of the X-ray and UV-only counterparts are summarized in Table \ref{tab:Table5} and Table \ref{tab:Table6}, respectively.  
 
Among the X-ray counterparts of $\delta$ Sct stars, two Algol-type binaries (RZ Cas and R CMa) are known to have $\delta$ Sct companions \citep{rod01}.  The X-ray origin of these systems are thought to be chromospheric coronal activity of the subgiant star \citep{van89,sin95}. Another two of them are known cataclysmic variable stars (CV), and were spectroscopically confirmed as binary system that consist of a white dwarf and $\delta$ Sct companion \citep{mcc99}. These two CVs have been studied by several authors (i.e., for J051523.8+324107 also known as 14 Aur or KW Aur, see \citealp{dan67,fit79,hod93}, for J212626.8+192224 also known as IK Peg, see \citealp{kur79, won94}). They also show EUV emission which is one of the properties of hot white dwarf stars \citep{buc95,mar97a,mar97b} and are selected as progenitors of type Ia supernovae \citep{par07}.
 
One of the X-ray counterparts is a previously known very fast rotating star, Altair ($\alpha$ Aql), which is the brightest $\delta$ Sct star in the sky \citep{buz05}. Even though Altair is inside the instability strip, no photometric variability was reported until the Wide Field Infrared Explorer (WIRE) satellite \citep{buz05}. The dominant source of X-ray emission is thought to be related to Altair's coronal activities \citep{rob09}. 
 
Although most of the X-ray counterparts do not have information about their X-ray origin, they are all interesting objects for further observations and studies because of their rare characteristics which show $\delta$ Sct pulsation and X-ray emission at the same time.  Especially five X-ray counterparts (J000910.1+590903, J034724.3+243513, J161441.0+335125, J195047.0+085159 and J120702.3-784428) seem very interesting objects because they also exhibit emission in optical and UV wavelength as well.\\

\begin{table*}
\begin{center}
 \caption{X-Ray, UV and EUV Catalogs \label{tab:Table4}}
 \begin{tabular}{cc}

 \tableline\tableline
ROSAT All-Sky Survey Bright Source Catalogue (1RXS) & \citet{vog99} \\
ROSAT All-Sky Survey Faint Source Catalogue (1RXS) & \citet{vog00} \\
BMW-Chandra source catalog (1BMC) & \citet{rom08} \\
XMM-Newton 2nd Incremental Source Catalogue (2XMMi) & \citet{wat09} \\
&\\
2RE Source Catalogue of the ROSAT Wide Field Camera & \\
All-Sky Survey of Extreme-Ultraviolet Sources (2RE) &  \citet{pye95} \\
&\\
&\\
Extreme-Ultraviolet Explorer Source Catalog (EUVE) & \citet{bow96} \\
& \citet{lam97} \\
& \citet{chr99}\\
&\\
Far Ultraviolet Spectroscopic Explorer  Observation Log (FUSE) & \citet{fus05} \\
Midcourse Space Experiment  Ultraviolet Point Source Catalog (MSX) & \citet{new06} \\

  \tableline
 \end{tabular}
\end{center}
\end{table*}

\section{Summary}
\label{sec:Summary}
 We compiled a new catalog of 1,578 $\delta$ Sct stars including the catalogs compiled by R2000, \citet{rod00b}, ASAS, ROSTE, TAOS and several individual findings published after R2000.  We highlight several key features and relationships between physical properties for the field and cluster member stars without companion objects.  Most of the properties are similar with those previously reported by other studies (e.g., R2000; \citealp{rod01}).  However, we also find indications of interesting correlations among pulsation and stellar parameters.  For example, the relations between full amplitude and period of the cluster member stars tell us that longer period $\delta$ Sct stars generally exhibit high-amplitude pulsations.  The final catalog was cross-matched with several X-ray and UV catalogs; 27 X-ray and 41 UV-only counterparts were found.  Among the X-ray and UV counterparts, there are two Algol-type eclipsing binaries, two cataclysmic variable stars and several binary candidates.  Further  observations will reveal the origin of X-ray/UV emission and the effect of binarity on the pulsation characteristics.  Our new catalog is accessible at the online website (\url{http://stardb.yonsei.ac.kr/DeltaScuti}).\\

\section*{Acknowledgements}
This research was supported by Basic Science Research Program through the National Research Foundation of Korea (NRF) funded by the Ministry of Education, Science and Technology (2012R1A1A2006924).  The analysis in this paper has been done using the Odyssey cluster supported by the FAS Research Computing Group at Harvard.  This research has made use of the SIMBAD database, operated at CDS, Strasbourg, France. This software uses source code created at the Centre de Donne'es astronomiques de Strasbourg (CDS), France.\\

\vspace{0.5cm}
{\small{

}}

\onecolumn
\begin{landscape}
\renewcommand{\arraystretch}{1.1}
{\small {
\begin{center}
\begin{longtable}{ccccccccccc}
\caption{Properties of 19 X-ray only sources and 8 X-ray/UV sources} \label{tab:Table5} \\

\tableline\tableline
\multicolumn{1}{c}{RA} &
\multicolumn{1}{c}{Dec} &
\multicolumn{1}{c}{Cross-matched ID} &
\multicolumn{1}{c}{Source Catalog$^a$} &
\multicolumn{1}{c}{$V$} &
\multicolumn{1}{c}{$B$} &
\multicolumn{1}{c}{Frequency} &
\multicolumn{1}{c}{$\Delta$$V$} &
\multicolumn{1}{c}{Spectral Type} &
\multicolumn{1}{c}{Type$^b$} &
\multicolumn{1}{c}{Designation$^c$} \\
\multicolumn{1}{c}{(hh:mm:ss)} & \multicolumn{1}{c}{(dd:mm:ss)} & &  & &  & \multicolumn{1}{c}{(cd$^{-1}$)} &  \multicolumn{1}{c}{(mmag)}  &  & & \\
\tableline
\endfirsthead

\multicolumn{11}{l}
{{\bfseries \tablename\ \thetable{} -- continued from previous page}} \\
\tableline\tableline
\multicolumn{1}{c}{RA} &
\multicolumn{1}{c}{Dec} &
\multicolumn{1}{c}{Cross-matched ID} &
\multicolumn{1}{c}{Source Catalog$^a$} &
\multicolumn{1}{c}{$V$} &
\multicolumn{1}{c}{$B$} &
\multicolumn{1}{c}{Frequency} &
\multicolumn{1}{c}{$\Delta$$V$} &
\multicolumn{1}{c}{SpType$^b$} &
\multicolumn{1}{c}{Type$^c$} &
\multicolumn{1}{c}{Designation$^d$} \\
\multicolumn{1}{c}{(hh:mm:ss)} & \multicolumn{1}{c}{(dd:mm:ss)} & &  & &  & \multicolumn{1}{c}{(cd$^{-1}$)}  &\multicolumn{1}{c}{(mmag)}&  & & \\
\tableline
\endhead

\tableline \multicolumn{11}{r}{{Continued on next page}}
\endfoot

\tableline
\endlastfoot
00:09:10 & +59:08:59 & J000910.1+590903 & ROSAT Bright & 2.28 & 2.66 & 9.911 & 33 & F2III & bm & $beta$ Cas \\
 &  & G117.5277-03.2775 & MSX &  &  &  &  &  &  &  \\
 &  & HD432 & FUSE &  &  &  &  &  &  &  \\
01:12:08 & +02:17:12 & J011207.8+021710 & 2XMMi & 11.899 & 12.569 & 5.663 & 160 &  & unknown &  \\
02:48:55 & +69:38:03 & J024854.7+693804 & ROSAT Bright & 6.26 & 6.411 & 62.112 & 20 & A3V & EA & RZ Cas \\
 &  & J024855.5+693803 & 2XMMi &  &  &  &  &  &  &  \\
03:44:31 & +32:06:22 & J034430.6+320628 & 2XMMi & 10.76 & 11.5 & 7.407 & 40 & F0m: & unknown & V0705 Per \\
 &  & 034431.2+320621 & BMW-Chandra &  &  &  &  &  &  &  \\
03:47:24 & +24:35:18 & J034724.3+243513 & ROSAT Bright & 7.673 & 7.905 & 16.584 & 20 & A4V & bm & V1228 Tau \\
 &  & J034724.0+243517 & 2XMMi &  &  &  &  &  &  &  \\
 &  & G166.2980-23.1107 & MSX &  &  &  &  &  &  &  \\
 &  & HD23628 & FUSE &  &  &  &  &  &  &  \\
04:28:39 & +15:52:15 & J042839.7+155217 & 2XMMi & 3.4 & 3.579 & 13.228 & 20 & A7III & bm & $\theta^2$ Tau \\
05:15:24 & +32:41:15 & J051523.8+324107 & ROSAT Bright & 5.01 & 5.232 & 11.351 & 80 & kA9hA9mF2 & WD & KW Aur \\
 &  & J0515+326 & EUVE &  &  &  &  &  &  &  \\
06:07:26 & -76:55:36 & J060725.1-765537 & ROSAT Bright & 9.63 & 10.222 & 4.998 & 230 & F7V & unknown &  \\
07:19:28 & -16:23:43 & J071928.0-162339 & ROSAT Bright & 5.7 & 6.046 & 21.277 & 10 & kA8hF1mF2 & EA & R CMa \\
 &  & G230.6704-01.4057 & MSX &  &  &  &  &  &  &  \\
07:58:30 & -60:37:46 & J075830.9-603746 & 2XMMi & 10.88 &  &  & 20 & A8V & unknown & V0419 Car \\
 &  & 075830.9-603748 & BMW-Chandra &  &  &  &  &  &  &  \\
07:58:33 & -60:49:26 & J075833.3-604925 & 2XMMi & 10.345 & 10.628 &  & 30 & A3V & unknown & V0420 Car \\
 &  & 075833.3-604926 & BMW-Chandra &  &  &  &  &  &  &  \\
08:39:09 & +19:35:33 & J083909.1+193530 & 2XMMi & 8.5 & 8.75 & 17.036 & 20 & A9V & unknown & BS Cnc \\
08:51:32 & +11:50:41 & 085132.1+115042 & BMW-Chandra & 12.25 & 12.52 & 18.832 & 20 & F0 & unknown & EW Cnc \\
12:07:05 & -78:44:28 & J120702.3-784428 & ROSAT Faint & 7.48 & 7.752 & 18.868 & 10 & A9III/IV & unknown & EF Cha \\
 &  & G300.6932-16.0626 & MSX &  &  &  &  &  &  &  \\
 &  & HD105234 & FUSE &  &  &  &  &  &  &  \\
12:49:08 & -41:12:26 & J124908.8-411225 & 2XMMi & 12.385 & 12.156 & 19.194 & 20 & A3V & unknown & V1041 Cen \\
13:26:28 & -47:31:02 & J132627.6-473102 & 2XMMi & 17.239 & 17.704 & 20.45 & 100 &  & unknown & $\omega$ Cen - NV319 \\
13:26:38 & -47:27:38 & 132638.3-472741 & BMW-Chandra & 17.096 &  & 20.833 & 80 &  & unknown & $\omega$ Cen - NV322 \\
13:26:40 & -47:29:11 & 132641.1-472911 & BMW-Chandra & 16.394 & 16.784 & 23.095 & 60 &  & unknown & $\omega$ Cen - NV312 \\
13:28:01 & -47:23:19 & J132801.5-472318 & 2XMMi & 9.411 & 10.001 &  & 30 & F8 & unknown & V1030 Cen \\
14:43:04 & -62:12:26 & J144304.5-621226 & 2XMMi & 7.4 & 7.559 & 28.249 & 10 & A3V & unknown & BT Cir \\
16:14:40 & +33:51:31 & J161441.0+335125 & ROSAT Bright & 5.23 & 5.829 & 0.877 & 50 & G1IV-V (k) & bm & TZ CrB \\
 &  & J161440.7+335129 & 2XMMi &  &  &  &  &  &  &  \\
 &  & 2RE J1614+335 & ROSAT-2RE &  &  &  &  &  &  &  \\
 &  & J1614+338 & EUVE &  &  &  &  &  &  &  \\
16:41:38 & +36:26:20 & J164138.2+362627 & 2XMMi & 17.12 &  & 15.314 & 250 &  & unknown & M13 -V47 \\
16:54:01 & -41:53:24 & J165402.2-415320 & 2XMMi & 14.01 &  & 15.625 & 40 &  & unknown & V1199 Sco \\
17:40:44 & -53:40:42 & 174044.1-534039 & BMW-Chandra & 15.34 & 15.68 & 26.178 & 40 &  & unknown & NGC 6397 - V11 \\
19:39:54 & -30:58:06 & J193953.4-305805 & 2XMMi & 17.09 & 17.45 & 24.39 & 29 &  & unknown & M55 - V27 \\
19:50:47 & +08:52:06 & J195047.0+085159 & ROSAT Bright & 0.76 & 0.981 & 15.773 & 4 & A7Vn & bm & $\alpha$ Aql \\
 &  & HD187642 & FUSE &  &  &  &  &  &  &  \\
21:26:26 & +19:22:32 & J212626.8+192224 & ROSAT Bright & 6.08 & 6.315 & 22.727 & 10 & kA6hA9mF0 & WD & IK Peg \\
 &  & HD204188 & FUSE &  &  &  &  &  &  &  \\
\end{longtable}
\end{center}

{\noindent}$^a$ ROSAT Bright : ROSAT All-Sky Survey Bright Source Catalog,
 ROSAT Faint : ROSAT All-Sky Survey Faint Source Catalog,
 2XMMi : The XMM-Newton 2nd Incremental Source Catalog,
 BMW-Chandra : The Brera Multi-scale Wavelet Chandra Survey,
 ROSAT-2RE : The ROSAT Wide Field Camera all-sky survey of extreme-ultraviolet sources,
 EUVE : Second Extreme Ultra-Violet Explorer Catalog,
 FUSE : Far Ultraviolet Spectroscopic Explorer,
 MSX :  The Midcourse Space Experiment Ultraviolet Point Source Catalog.\\
$^b$ Spectral type. \\
$^c$ WD : white dwarf companion, EA : Algol-type variable star, bm : binary or multiple star, unknown : no specific information.\\
$^d$ GCVS Designation.
}}
\end{landscape}
\twocolumn

\onecolumn
\begin{landscape}
\renewcommand{\arraystretch}{1.1}
{\small {
\begin{center}
\begin{longtable}{ccccccccccc}
\caption{Properties of 41 UV-only sources} \label{tab:Table6} \\

\tableline\tableline
\multicolumn{1}{c}{RA} &
\multicolumn{1}{c}{Dec} &
\multicolumn{1}{c}{Cross-matched ID} &
\multicolumn{1}{c}{Source Catalog$^a$} &
\multicolumn{1}{c}{$V$} &
\multicolumn{1}{c}{$B$} &
\multicolumn{1}{c}{Frequency} &
\multicolumn{1}{c}{$\Delta$$V$} &
\multicolumn{1}{c}{Spectral Type} &
\multicolumn{1}{c}{Type$^b$} &
\multicolumn{1}{c}{Designation$^c$} \\
\multicolumn{1}{c}{(hh:mm:ss)} & \multicolumn{1}{c}{(dd:mm:ss)} & &  & &  & \multicolumn{1}{c}{(cd$^{-1}$)} &  \multicolumn{1}{c}{(mmag)}  &  & & \\
\tableline
\endfirsthead

\multicolumn{11}{l}%
{{\bfseries \tablename\ \thetable{} -- continued from previous page}} \\
\tableline\tableline
\multicolumn{1}{c}{RA} &
\multicolumn{1}{c}{Dec} &
\multicolumn{1}{c}{Cross-matched ID} &
\multicolumn{1}{c}{Source Catalog$^a$} &
\multicolumn{1}{c}{$V$} &
\multicolumn{1}{c}{$B$} &
\multicolumn{1}{c}{Frequency} &
\multicolumn{1}{c}{$\Delta$$V$} &
\multicolumn{1}{c}{SpType$^b$} &
\multicolumn{1}{c}{Type$^c$} &
\multicolumn{1}{c}{Designation$^d$} \\
\multicolumn{1}{c}{(hh:mm:ss)} & \multicolumn{1}{c}{(dd:mm:ss)} & &  & &  & \multicolumn{1}{c}{(cd$^{-1}$)}  &\multicolumn{1}{c}{(mmag)}&  & & \\
\tableline
\endhead

\tableline \multicolumn{11}{r}{{Continued on next page}}
\endfoot

\tableline
\endlastfoot
00:43:48 & +42:16:56 & G121.4228-20.5669 & MSX & 9.36 & 9.718 & 8.006 & 240 & F3IV-V & bm & CC And \\
00:50:41 & -50:59:13 & G303.2241-66.1409 & MSX & 5.24 & 5.596 & 5.447 & 40 & F3III & unknown & $\rho$ Phe \\
02:10:25 & +59:58:47 & G132.6663-01.3911 & MSX & 6.66 & 6.994 & 9.158 & 48 & kF0hF2mF2(III) & unknown & V0784 Cas \\
03:47:03 & +24:49:12 & G166.0675-22.9921 & MSX & 8.28 & 8.65 & 31.25 & 18 & F0V & bm & V0534 Tau \\
04:16:39 & -64:18:56 & 041639-641851 & MSX04 & 7.97 & 8.266 & 14.925 & 20 & F0III & unknown & TX Ret \\
05:38:05 & -01:15:22 & G205.4934-16.8532 & MSX & 9.92 & 10.24 & 10.309 & 20 & F0Ve & unknown & V1247 Ori \\
06:14:08 & +23:59:11 & G187.4691+03.0744 & MSX & 7.53 & 7.885 & 5.316 & 50 & F0 & bm & PV Gem \\
06:19:37 & +59:00:39 & G155.5948+19.1163 & MSX & 4.44 & 4.472 & 15.385 & 300 & A2V & bm & UZ Lyn \\
07:07:56 & -04:40:40 & G218.9778+01.5291 & MSX & 6.92 & 7.217 & 4.32 & 20 & A9V & bm & V0752 Mon \\
07:11:23 & -00:18:07 & G215.4813+04.3013 & MSX & 5.44 & 5.75 & 10.0 & 50 & F1IV & unknown & V0571 Mon \\
07:27:08 & -17:51:53 & G232.8306-00.4863 & MSX & 5.6 & 5.914 & 6.017 & 20 & A9Vn & bm & NR CMa \\
08:27:36 & -53:05:19 & G269.2568-08.4423 & MSX & 5.08 & 5.336 & 14.286 & 10 & F0III-IV & bm & GU Vel \\
08:58:52 & -47:14:05 & G267.7083-00.9048 & MSX & 5.17 & 5.438 & 15.385 & 20 & F0V & bm & FZ Vel \\
09:11:07 & -43:16:11 & G266.2026+03.3494 & MSX & 7.88 & 8.194 & 4.292 & 20 & A8-9III & unknown & MP Vel \\
09:56:54 & -27:28:31 & G261.8969+21.1682 & MSX & 6.32 & 6.493 &  & 10 & A4V? & unknown & BF Ant \\
10:05:01 & -56:53:53 & G281.6679-01.0608 & MSX & 6.86 & 7.23 & 8.271 & 30 & F2III/IV & unknown & V0336 Vel \\
10:05:13 & -79:03:44 & 100512-790341 & MSX04 & 7.32 & 7.568 &  & 80 & A3/5III/IV & unknown & ER Cha \\
10:13:22 & -51:13:59 & G279.3845+04.2657 & MSX & 5.27 & 5.527 & 7.981 & 20 & F0Vn & unknown & LW Vel \\
11:49:03 & +14:34:19 & HD102647 & FUSE & 2.14 & 2.23 &  & 25 & A3V & bm & $\beta$ Leo \\
12:02:06 & +43:02:44 & G151.8990+71.2009 & MSX & 5.22 & 5.503 & 25.0 & 20 & kA4hA6mA7 & unknown & DP UMa \\
12:23:47 & +42:32:34 & G141.2302+73.5894 & MSX & 6.03 & 6.396 & 8.598 & 70 & F3IV & unknown & AI CVn \\
13:12:49 & -61:32:42 & G305.4777+01.2215 & MSX & 7.26 & 7.605 & 9.141 & 50 & A9/F0III/IV & unknown & V0954 Cen \\
14:29:58 & -56:07:52 & G316.3983+04.1416 & MSX & 6.97 & 7.203 & 18.939 & 30 & A6V & bm & V0853 Cen \\
15:01:02 & -64:34:34 & G316.2853-05.1186 & MSX & 6.56 & 6.858 & 6.329 & 70 & A9/F0IV/V & unknown & BV Cir \\
15:18:34 & +02:05:00 & NGC5904-ZNG1 & FUSE & 15.3 &  & 11.136 & 100 &  & unknown &  \\
15:42:44 & +26:17:44 & HD140436 & FUSE & 3.81 & 3.83 & 33.333 & 60 & A0V & bm & $\gamma$ CrB \\
16:27:51 & -49:07:36 & G334.9455-00.1925 & MSX & 13.51 &  & 16.807 & 3 &  & unknown &  \\
17:32:24 & -34:16:46 & G353.9809-00.4715 & MSX & 6.16 & 6.52 & 4.606 & 30 & F2V & unknown & V0949 Sco \\
17:42:16 & -32:31:24 & G356.5669-01.2673 & MSX & 7.85 & 8.25 & 6.667 & 500 & F0V & unknown & V0703 Sco \\
18:42:16 & -09:03:09 & G023.8330-02.0981 & MSX & 4.7 & 5.058 & 5.16 & 190 & F2II-III & bm & $\delta$ Sct \\
18:59:51 & +11:26:42 & G044.0788+03.3517 & MSX & 7.72 & 8.183 & 6.2 & 56 & F0 & unknown & V1438 Aql \\
19:19:39 & +12:22:29 & G047.1398-00.5184 & MSX & 5.53 & 5.795 & 6.849 & 13 & A9III & bm & V1208 Aql \\
19:42:49 & +29:19:54 & G064.6016+02.9162 & MSX & 6.54 & 6.877 & 11.364 & 20 & F1III & unknown & V1276 Cyg \\
20:10:33 & +26:54:14 & G065.7093-03.5691 & MSX & 5.51 & 5.597 & 8.23 & 10 & A2IV & unknown &  \\
20:10:45 & +26:44:36 & 201045+264437 & MSX04 & 10.23 & 10.67 & 17.857 & 30 & A7IV & unknown & V0381 Vul \\
20:14:14 & +28:41:41 & G067.6569-03.2692 & MSX & 5.19 & 5.381 & 5.316 & 16 & A5Vn & unknown & NU Vul \\
20:56:35 & -05:42:06 & 205635-054201 & MSX04 & 13.285 &  & 12.903 & 450 &  & unknown &  \\
21:14:47 & +38:02:43 & G082.8535-07.4321 & MSX & 3.74 & 4.133 & 12.048 & 20 & F2+ V & bm & $\tau$ Cyg \\
22:15:02 & +57:02:37 & G102.8657+00.3947 & MSX & 4.18 & 4.458 & 24.272 & 20 & F0V(Sr) & bm & $\epsilon$ Cep \\
22:48:30 & -10:33:20 & G056.7989-56.6757 & MSX & 6.19 & 6.464 & 11.494 & 20 & F1V & unknown & FM Aqr \\
23:28:25 & -25:25:14 & G033.4840-71.3326 & MSX & 6.9 & 7.06 & 8.482 & 20 & A3 & unknown & BS Scl \\
\end{longtable}
\end{center}

{\noindent}$^a$ ROSAT-2RE : The ROSAT Wide Field Camera all-sky survey of extreme-ultraviolet sources, EUVE : Second Extreme Ultra-Violet Explorer Catalog, FUSE : Far Ultraviolet Spectroscopic Explorer, MSX :  The Midcourse Space Experiment Ultraviolet Point Source Catalog, MSX04 : MSX UV Point Source Catalog.\\
$^b$ Spectral type. \\
$^c$ bm : binary or multiple star, unknown : no specific information.\\
$^d$ GCVS Designation.
}}
\end{landscape}
\twocolumn

\label{lastpage}
\end{document}